\documentclass[
    10pt,
    twocolumn,
    superscriptaddress,
    prb,
    showpacs
]{revtex4}

\usepackage{graphicx}
\usepackage{amssymb}
\usepackage{amsmath}
\usepackage{color}
\usepackage{footmisc}

\newcommand{\Eqref}[1]{Eq.~(\ref{#1})}
\newcommand{\Figref}[1]{Fig.~\ref{#1}}
\newcommand{\Tabref}[1]{Tab.~\ref{#1}}
\newcommand{\Secref}[1]{Sec.~\ref{#1}}

\begin{document}
\title{From tunneling to contact: Inelastic signals in an atomic gold junction}


\author{Thomas \surname{Frederiksen}}
\email{thf@mic.dtu.dk} \affiliation{MIC -- Department of Micro and
Nanotechnology, NanoDTU, Technical University of Denmark,\\
{\O}rsteds Plads, Bldg.~345E, DK-2800 Lyngby, Denmark}
\author{Nicol\'{a}s \surname{Lorente}}
\affiliation{Laboratorie Collisions, Agr\'{e}gats,
R\'{e}activit\'{e}, IRSAMC, Universit\'{e} Paul Sabatier,\\
118 Route de Narbonne, F-31062 Toulouse, France}
\author{Magnus \surname{Paulsson}}
\author{Mads \surname{Brandbyge}}
\affiliation{MIC -- Department of Micro and Nanotechnology, NanoDTU,
Technical University of Denmark,\\ {\O}rsteds Plads, Bldg.~345E,
DK-2800 Lyngby, Denmark}

\date{\today}


\begin{abstract}
The evolution of electron conductance in the presence of inelastic
effects is studied as an atomic gold contact is formed evolving from
a low-conductance regime (tunneling) to a high-conductance regime
(contact). In order to characterize each regime, we perform density
functional theory (DFT) calculations to study the geometric and
electronic structures, together with the strength of the atomic
bonds and the associated vibrational frequencies. The conductance is
calculated by first evaluating the transmission of electrons through
the system, and secondly by calculating the conductance change due
to the excitation of vibrations. As found in previous studies
[Paulsson \textit{et al.}, Phys. Rev. B. {\bf 72}, 201101(R) (2005)]
the change in conductance due to inelastic effects permits to
characterize the crossover from tunneling to contact. The most
notorious effect being the crossover from an increase in conductance
in the tunneling regime to a decrease in conductance in the contact
regime when the bias voltage matches a vibrational threshold. Our
DFT-based calculations actually show that the effect of vibrational
modes in electron conductance is rather complex, in particular when
modes localized in the contact region are permitted to extend into
the electrodes. As an example, we find that certain modes can give
rise to decreases in conductance when in the tunneling regime,
opposite to the above mentioned result. Whereas details in the
inelastic spectrum depend on the size of the vibrational region, we
show that the overall change in conductance is quantitatively well
approximated by the simplest calculation where only the apex atoms
are allowed to vibrate. Our study is completed by the application of
a simplified model where the relevant parameters are obtained from
the above DFT-based calculations.
\end{abstract}

\maketitle


\section{Introduction}

Recent experimental advances have permitted to probe electron
transport processes at the atomic scale.\cite{Agrait:2003} Junctions
can be formed that support current flow through atom-sized
constrictions or even single molecules. Atomic vibrations become
detectable and very dependable on the environmental temperature.
According to the distance between electrodes, the conductance can
vary several orders of magnitude when the applied voltages are
small, typically below the eV scale. This behavior is due to the
exponential dependence of current with distance when the conductance
is due to an electron tunneling process. However, at short electrode
distances, the current levels off and saturates: the contact regime
is reached. The conductance is maximum in this case and a
high-conductance regime is attained. The physics in these two
regimes can be very different.

The low-conductance regime has been thoroughly studied with the
scanning tunneling microscope (STM). The initial inelastic effects
were realized by showing the increase in conductance on an acetylene
molecule when the bias voltage matched the C--H stretch
mode.\cite{Stipe:1998} The proof that the mode was indeed excited
was the isotopical effect that the changes of conductance showed
when replacing C$_2$H$_2$ by C$_2$D$_2$. This finding paved the way
to vibrational spectroscopy with sub-{\AA}ngstr\"om spatial
resolution, permitting the identification of chemical components of
matter on the atomic scale.\cite{Pascual:2005,Komeda:2005} The first
experimental evidence of mode excitation in the high-conductance
regime was achieved in monatomic gold wires.\cite{Agrait:2002} The
conductance of the wires showed clear reductions at thresholds that
were proven to originate in the backscattering of electrons from
some selected vibrations of the
wires.\cite{Agrait:2002,Frederiksen:2004} Similarly, experiments
with the break junction geometry have also revealed signatures in
the conductance related to several vibrational modes of a single
H$_2$ molecule trapped between the electrodes.\cite{Smit:2002}

The emerging picture is that in the tunneling or low-conductance
regime, the excitation of vibrations leads to increases in
conductance at the corresponding voltage thresholds, while in the
contact or high-conductance regime, the effect of vibrations is to
reduce the conductance. Theoretical studies in the weak
electron-vibration coupling regime have shown that the lowest order
expansion\cite{Galperin:2004} is capable of correlating this
behavior with a single parameter: the eigenchannel transmission
probability $\tau$.\cite{Paulsson:2005,Viljas:2005,delaVega:2006} In
the simplified case of a single electronic level connected with two
electrodes under symmetrical conditions, the inelastic effects (of a
vibrationally mediated on-site modulation) go from increases in the
conductance for $\tau < 1/2$ to decreases for $\tau > 1/2$. In this
way, the behavior of the inelastic conductance would define the
crossover from tunneling to contact. There is experimental evidence
showing that this picture is indeed more complex. The excitation of
the O--O stretch mode of the chemisorbed O$_2$ molecule on
Ag(110)~\cite{Hahn:2000} leads to a decrease of the tunneling
current (instead of an increase) in opposition with most cases in
the low-conductance regime.\cite{Persson:1984,Lorente:2004}

The aim of the present work is to analyze the continuous evolution
from tunneling to contact in a model system constituted by a
junction of gold atoms, which provides an almost perfect realization
of a single transmission channel system. The definition of when a
given atomic system correspond to one of the two cases analyzed
above is already problematic, hence we address this issue by
investigating the behavior of different properties of the junction
with the interatomic distance. Initially, we are interested in
studying the crossover from tunneling to contact by evaluating the
total energy, the strain, and the modification of vibrational modes
as the electrode distance decreases. This allows us to find a range
of distances where the junction behaves as either two independent
systems or a strongly coupled one. The second part of this work
evaluates the effect of the interatomic distance in electron
transmission; this allows us to study the elastic conductance within
Landauer's formalism. The correlation of the transmission against
the interatomic relaxation permits a clear identification of both
regimes as well as the transition region. Finally, the inelastic
properties of the conductance are studied in the different regimes.
The inelastic signals are interpreted in a simplified model that
captures the calculated behavior and illustrates the fundamental
concepts.

The continuous transition from tunneling to contact is
experimentally challenging, since most metallic
point contacts (including Au) usually exhibits a sudden jump in the
conductance when the surfaces are brought into
contact.\cite{Untiedt:2006} On the other hand, experiments with a
low-temperature STM on Cu(111) and Ag(111) surfaces have shown that
both sharp jumps as well as smooth variations can be obtained in the
crossover from tunneling to contact: when the tip is approached over
a clean surface one observes a jump in conductance (related to the
transfer of the tip-atom to the surface), whereas over an isolated
metallic adatom the evolution is smooth and
reversible.\cite{Limot:2005}  To our knowledge there is no
measurement of the evolution of the inelastic signals in the formation of
a metallic point contact, likely owing the relative weak effect
(conductance changes expected to be less than 1\%) and difficulties
with noise. Despite this we envision that our idealized model system
is not unrealistic, and can provide a useful framework for
investigating the complicated interplay between chemical bonding,
electron conduction, atomic vibrations, etc. Our first-principles
treatment further addresses all of these issues in a unified way to
provide quantitative predictions.

\begin{figure}[t!]
\centering
\includegraphics[width=.6\columnwidth]{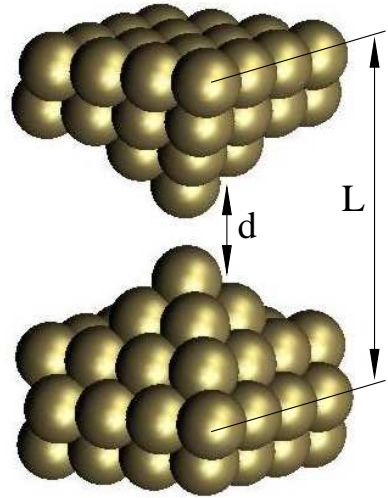}
\caption{(Color online) Generic setup for the calculation of
structural properties of the atomic gold junction. The periodic
supercell consists of a $4\times 4$ representation of two Au(100)
surfaces sandwiching two pyramids pointing towards each other. The
characteristic electrode separation $L$ is measured between the
second-topmost surface layers, since the surface layer itself is
relaxed and hence deviates on the decimals from the bulk values. The
interatomic distance between the apex atoms is denoted $d$.}
  \label{fig:Tunnel2Contact}
\end{figure}


\section{Theory}

The present work can be divided by the different methods that we
have used. In order to study the structural properties of the atomic
junction the standard DFT \textsc{Siesta}\cite{Soler:2002} method is
used. The elastic conductance is evaluated from the transmission
function of the atomic junction calculated with
\textsc{TranSiesta},\cite{Brandbyge:2002} and the inelastic
contribution is performed using the method presented in
Ref.~[\onlinecite{Paulsson:2005,Frederiksen:2006}].

The system representing the atomic junction is depicted in
\Figref{fig:Tunnel2Contact}. We consider a periodic supercell with a
$4\times 4$ representation of two Au(100) surfaces sandwiching two
pyramids pointing towards each other. The characteristic electrode
separation $L$ will be measured between the second-topmost surface
layers, since the surface layer itself is relaxed and hence deviates
on the decimals from the bulk values. The corresponding calculations
with the \textsc{Siesta} method are carried out using a single zeta
plus polarization (SZP) basis  with a confining energy of 0.01 Ry
(corresponding to the 5$d$ and 6$(s,p)$ states of a free Au atom),
the generalized gradient approximation (GGA) for the
exchange-correlation functional, a cutoff energy of 200 Ry for the
real-space grid integrations, and the $\Gamma$-point approximation
for the sampling of the three-dimensional Brillouin zone. The
interaction between the valence electrons and the ionic cores are
described by a standard norm-conserving Troullier-Martins
pseudo-potential generated from a relativistic atomic calculation.

The calculations of the vibrations are performed by diagonalization
of the dynamical matrix extracted from finite differences (with
corrections for the egg-box effect, i.e.,~the movement of basis
orbitals---following the displaced atoms---with respect to the real
space integration grid).\cite{Frederiksen:2006} As the active atoms
we consider initially---for pedagogical purposes---just the two apex
atoms and compare afterwards the results when the vibrational region
is enlarged.

The transport calculation naturally considers infinite electrodes by
including the DFT self-energy calculated for infinite atomistic
leads in the conduction equations.\cite{Brandbyge:2002} Since we are
here interested in the low-bias regime (of the order of the
vibrational frequencies) it suffices to calculate the electronic
structure in equilibrium in order to describe the elastic transport
properties. While the transmission generally involves a sampling
over $\mathbf{k}$-points we here approximate it with its
$\Gamma$-point value; this has previously been shown to be a
reasonable approximation for supercells of similar dimensions in the
case of atomic gold wires.\cite{Frederiksen:2006}

Finally, the inelastic transport calculations are performed using
the nonequilibrium Green's function (NEGF) formalism combined with
the electrode couplings $\mathbf{\Gamma}_{L,R}$ extracted from the
\textsc{TranSiesta} calculations and the electron-vibration
couplings $\mathbf{M}^\lambda$ (corresponding to modes $\lambda$
with energies $\hbar\omega_\lambda$) from the finite-difference
method.\cite{Frederiksen:2006} According to the lowest order
expansion (LOE)\cite{Paulsson:2005,Viljas:2005} the inelastic
current reads
\begin{widetext}
\begin{eqnarray}
\label{eq:LOEcurrent}
 I^\mathrm{LOE} &=& \textrm{G}_0 V \mathrm{Tr}
[\mathbf{G} \mathbf{\Gamma}_R \mathbf{G}^\dag \mathbf{\Gamma}_L]
\nonumber \\ && + \sum_\lambda {\cal I}^\mathrm{sym}_\lambda(V,T,
\langle n_\lambda\rangle) \,
   \mathrm{Tr}\Big[
     \mathbf{G}^\dag \mathbf{\Gamma}_L \mathbf{G} \Big\{
     \mathbf{M}^\lambda \mathbf{A}_R  \mathbf{M}^\lambda +
       \frac{i}{2}( \mathbf{\Gamma}_R \mathbf{G}^\dag \mathbf{M}^\lambda \mathbf{A} \mathbf{M}^\lambda - \mathrm{h.c.})
       \Big\}
  \Big]  \nonumber \\
 & & +
\sum_\lambda{\cal I}^\mathrm{asym}_\lambda(V, T)\,
  \mathrm{Tr}\Big[
         \mathbf{G}^\dag \mathbf{\Gamma}_L \mathbf{G} \Big\{
           \mathbf{\Gamma}_R \mathbf{G}^\dag \mathbf{M}^\lambda
           ( \mathbf{A}_R-\mathbf{A}_L ) \mathbf{M}^\lambda +
           \mathrm{h.c.}
          \Big\}
    \Big] ,
 \label{eq:current1} \\
{\cal I }^\mathrm{sym}_\lambda & = &
      \frac{e }{\pi \hbar} \left( {2 e V} \langle n_\lambda \rangle +
      \frac{\hbar \omega_\lambda-{e V}}{e^{\beta(\hbar \omega_\lambda-e V)}-1}-
      \frac{{\hbar \omega_\lambda}+{e V}}{e^{\beta(\hbar \omega_\lambda+e V)}-1}\right) ,
\label{eq:currentNormal} \\
 {\cal I}^{\mathrm{asym}}_\lambda&=&
    \frac{e}{\hbar}\int_{-\infty}^{\infty} \frac{d\varepsilon}{2\pi} \left[n_\textrm{F}(\varepsilon) -n_\textrm{F}(\varepsilon-e V)\right] \,
    {\cal H}_{\varepsilon'} \{{n_\textrm{F}(\varepsilon'+\hbar \omega_\lambda)
    -n_\textrm{F}(\varepsilon'-\hbar \omega_\lambda)}\}(\varepsilon) ,
   \label{eq:currentHilbert}
\end{eqnarray}
where G$_0=2e^2/$h is the conductance quantum, $V$ the external bias
voltage, $\langle n_\lambda \rangle$ the occupation of mode
$\lambda$, $n_\textrm{F}(\varepsilon)$ the Fermi function, ${\cal
H}$ the Hilbert transform, and $\beta=1/\textrm{k}_BT$ the inverse
temperature. The retarded Green's function $\mathbf{G}$, the
spectral function $\mathbf{A}=i(\mathbf{G}-\mathbf{G}^\dag)$, as
well as the electrode couplings $\mathbf{\Gamma}_{L,R}$ are all
evaluated at the Fermi energy in the LOE scheme. For convenience we
have also defined the quantities
$\mathbf{A}_{L,R}=\mathbf{G}\mathbf{\Gamma}_{L,R}\mathbf{G}^\dag$
such that $\mathbf{A}=\mathbf{A}_L+\mathbf{A}_R$. The sums in
\Eqref{eq:LOEcurrent} runs over all modes $\lambda$ in the
vibrational region. For a symmetric system (such as the present one
for the atomic junction) it can be shown that the asymmetric terms
in the current expression vanishes. Furthermore, at low temperatures
($\beta\rightarrow\infty$) and in the externally damped limit
($\langle n_\lambda \rangle\approx 0$) the inelastic conductance
change from each mode $\lambda$ (beyond the threshold voltage
$eV>\hbar\omega_\lambda$) is given by
\begin{eqnarray}
\label{eq:Conductance} \delta G_\lambda &=&
\mathrm{G}_0\,\mathrm{Tr}\Big[\mathbf{G}^\dag \mathbf{\Gamma}_L
\mathbf{G} \Big\{ \mathbf{M}^\lambda \mathbf{G} \mathbf{\Gamma}_R
\mathbf{G}^\dag \mathbf{M}^\lambda +\frac{i}{2} (\mathbf{\Gamma}_R
\mathbf{G}^\dag \mathbf{M}^\lambda \mathbf{A} \mathbf{M}^\lambda -
\mathrm{h.c.})\Big\}\Big].
\end{eqnarray}
\end{widetext}
From this expression we note that $\delta G_\lambda$ can be either
positive or negative, depending on sign of the trace.


\section{Structural and vibrational properties of the atomic junction}

\begin{figure}[t!]
  \centering
  \includegraphics[width=.9\columnwidth]{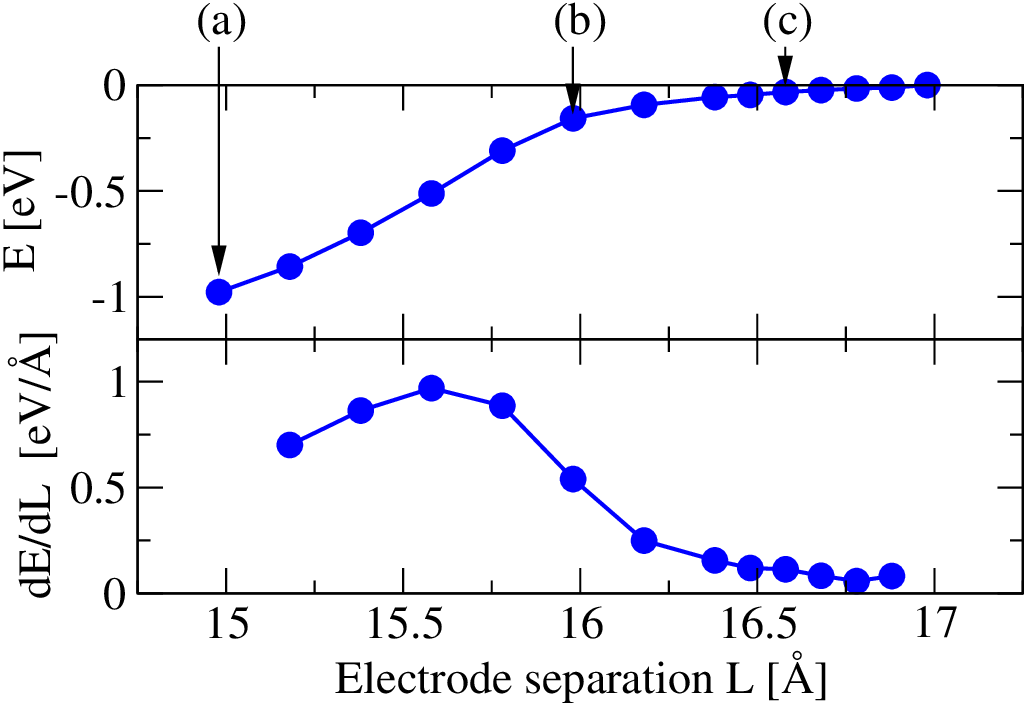}
  \caption{(Color online) Total energy differences and the numerical derivatives as a function of the
  electrode separation. The lower part of the figure describes the
  strain on the unit cell along the transport direction. The onset
  of chemical interactions is clearly seen around $L=16.0$ {\AA} where the force
  experience a significant increase.
  (a), (b), and (c) are three representative electrode separations
  of the three regimes considered in this article.}
  \label{fig:EtotVsLength}
\end{figure}

As the electrode separation is decreased, we relax in each step the
apex atoms, the base atoms of the pyramids, and the first-layer
atoms until residual forces are smaller than 0.02 eV/{\AA}. This
allows us to obtain the evolution of the (Kohn-Sham) total energy
$E$ of the system as a function of the electrode distance,
see~\Figref{fig:EtotVsLength}. We find that the energy is reduced
(of the order 1 eV) by the attractive interaction between the apex
atoms, due to the formation of a covalent bond at short distances,
\Figref{fig:EtotVsLength}(a). The slope of the energy presents a
rapid change for distances shorter than $L=16.0$ {\AA}. This is more
clearly seen in the lower part of \Figref{fig:EtotVsLength} where
the strain---or force on the unit cell---is represented. This force
is evaluated as the numerical derivative of the total energy with
respect to electrode separation. Here, the onset of chemical
interactions is clearly seen around $L=16.0$ {\AA},
\Figref{fig:EtotVsLength}(b), where the force experiences a
significant increase reaching a maximum at $L=15.6$ {\AA}.

\begin{figure}[t!]
  \centering
  \includegraphics[width=.9\columnwidth]{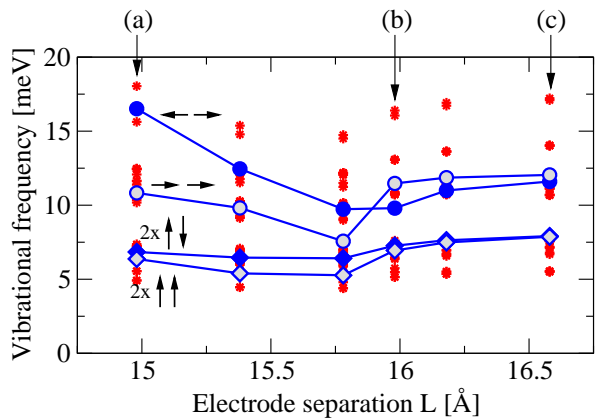}
  \caption{(Color online) Vibrational frequencies versus electrode displacement.
  The connected data series refer to the situation where only
  the two apex atoms are vibrating (resulting in the six vibrational modes
  indicated in the plot); circles symbolize the two longitudinal modes
  (CM and ABL) and diamonds the four (pair-wise degenerate)
  transversal modes. The red stars are the corresponding vibrational
  frequencies when also the pyramid bases are considered active.
  The three regimes are clearly identifiable:
  (a) concerted apex vibrations, (b) crossover
  where the stretch modes become degenerate, and (c)
  independent apex vibrations.}
  \label{fig:FreqVsLength}
\end{figure}

The evolution of the interaction between the apex atoms with
distance is also revealed in the study of the vibrational modes.
This is presented in \Figref{fig:FreqVsLength}, where the blue
connected data points correspond to the 6 modes where only the apex
atoms vibrate, and the red crosses to the 30 modes where also the
pyramid bases vibrate. These modes follow different behavior with
the electrode separation.

In the following we analyze the simplest case of just the two apex
atoms. Generally, two longitudinal stretch modes (represented with
connected circles in \Figref{fig:FreqVsLength}) line up the highest
in energy. For an electrode distance larger than $L=16.5$ {\AA}
these correspond to the isolated (i.e., decoupled and hence
degenerate) stretch modes of each apex atom,
\Figref{fig:FreqVsLength}(c). As the electrodes are approached, the
attractive apex-apex interaction leads to a slight displacement of
the apex atoms away from the base of the pyramids. The consequence
is a small weakening of the apex-atom coupling to the base which
results in decreasing frequencies, i.e., to softening of the modes.
Another consequence of the increasing interaction is the splitting
of the degenerate modes into a symmetric (out--of--phase) and an
antisymmetric (in--phase) mode. We will refer to these as the
alternating bond length (ABL) mode\cite{Frederiksen:2004} and the
center of mass (CM) mode, respectively. When the electrode
separation reaches the region between $L=15.8$ {\AA} and $L=16.0$
{\AA} the frequencies drop significantly,
\Figref{fig:FreqVsLength}(b). This points again at the chemical
interaction crossover that we presented in the previous paragraph:
now the interaction between the apex atoms becomes comparable with
the interaction with the electrodes and hence weakens the stretch
modes initially set by the interaction between the apex atom with
the base of the pyramid. As the apex-apex interaction grows larger,
the modes start to increase in frequency and further show an
significant split, \Figref{fig:FreqVsLength}(a).

The behavior of the two stretch modes of \Figref{fig:FreqVsLength}
is easily understood with a simple one-dimensional elastic model of
two masses, each coupled to infinite-mass system with a spring
constant $k_1$, and interconnected by another spring constant $k_2$.
The frequencies of the two stretch modes are then
$\omega_\textrm{CM}=\sqrt{k_1/m}$ (in--phase) and
$\omega_\textrm{ABL}=\sqrt{(k_1 + 2 k_2)/m}$ (out--of--phase), where
$m$ is the mass of each atom. Note that in the tunneling regime the
apex-apex interaction is attractive, cf.~\Figref{fig:EtotVsLength},
which would correspond to a negative value of $k_2$. When the bond
has been formed, $k_2$ can be represented as classical (positive)
spring constant. This model essentially captures the evolution of
the stretch modes. In particular, the sign change of $k_2$ at the
chemical instability explains the mode crossing between $L=15.8$
{\AA} and $L=16.0$ {\AA}, \Figref{fig:FreqVsLength}(b),
 and why the CM mode has a higher frequency
than the ABL mode in the tunneling case, and vice versa in the
contact case.

The analysis of the modes with electrode distance thus permits us to
recover the same range of distances with the chemical crossover as
in the preceding section concerning the total energy and strain.
This identification is also possible from the more realistic
calculation that includes the vibration of the base atoms.


\section{Elastic conductance}

\begin{figure}[t!]
  \centering
  \includegraphics[width=\columnwidth]{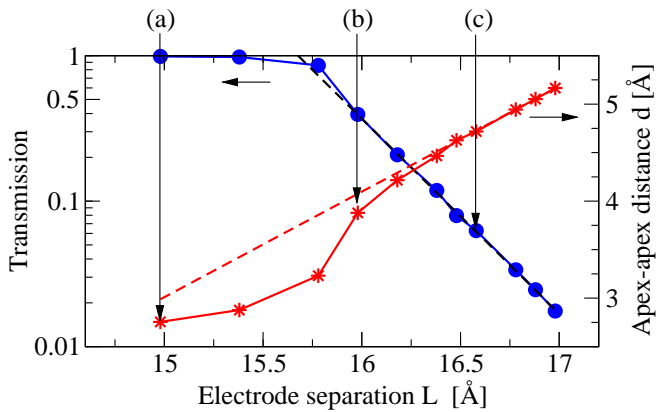}
  \caption{(Color online) Transmission $\tau$ (blue
  disks) and apex-apex distance $d$ (red crosses) versus electrode separation $L$.
  In the tunneling regime the transmission decays exponentially
  with separation as indicated with the dashed line (corresponding
  to a tunneling parameter $\Lambda=1.54~\textrm{\AA}^{-1}$). The point
  at (a) corresponds well with the contact region of transmission one and
  closest apex separation, (b) is near half transmission and the instability
  in apex separation, (c) is finally the tunneling regime, where the apex atoms
  are independent.}
  \label{fig:TransVsApexDist}
\end{figure}

In this study we are interested in the low-bias regime. Hence the
elastic conductance is determined via Landauer's formula by the
transmission at the Fermi energy $\varepsilon_\textrm{F}$. As
expected for the gold contact, we find that the total transmission
is essentially due to a single eigenchannel (for the geometries
considered here the contribution from the secondary channel is at
least three orders of magnitude smaller). Figure
\ref{fig:TransVsApexDist} plots the transmission $\tau$ and the
apex-apex distance $d$ as a function of electrode separation $L$. In
the tunneling regime the transmission is characterized by an
exponential decay with separation. It is instructive to compare this
with the transmission probability $T\propto \exp(-2\Lambda D)$ for a
rectangular one-dimensional barrier, where $\Lambda= \sqrt{2m_e
\Phi}/\hbar$ is a characteristic tunneling length, $\Phi$ the
apparent barrier height, and $D$ the barrier width (valid for
$\Lambda D\gg 1$). As shown in \Figref{fig:TransVsApexDist}, an
exponential fit to the calculated tunneling data leads to a
tunneling parameter $\Lambda=1.54~\textrm{\AA}^{-1}$ which would
correspond to an apparent barrier height of the order $\Phi\approx
9.1$ eV. Compared with measurements of the work function on
perfectly flat Au surfaces (5.31--5.47 eV)\cite{Handbook:2007} this
value is certainly high. On the other hand $\Phi$ is not very well
determined from an exponential fit to data spanning only one decade.
Another contribution to a relatively large barrier height could be
geometric effects from the pyramidal shapes.

The deviation from the
exponential tunneling behavior (visible around $L=16.0$ {\AA}) is a
clear indication of the crossover to contact. The contact regime is
characterized by a constant transmission equal to unity since an
atomic gold junction has effectively only one conduction channel.
The value $\tau=1/2$ to define the crossover between contact and
tunneling is somewhat arbitrary, however a detailed comparison with
the previous section justifies this definition. Indeed,
\Figref{fig:TransVsApexDist} also shows the behavior of the
apex-atom distance $d$ with electrode separation, permitting us to
make contact with the chemical crossover defined previously. Between
$L=15.8$ {\AA}~and 16.0~{\AA}~we find that the apex-apex distance
has changed by almost 0.7 {\AA}. This shows that at these electrode
distances, there is an instability that drive the formation of a
covalent bond between apex atoms. This agrees with the conclusion
from both total energy, strain and frequency calculations that the
crossover takes place between 15.8 {\AA}~and 16.0~{\AA}. From
\Figref{fig:TransVsApexDist} we identify a transmission 1/2
associated with $L=15.9$ {\AA} ($d=3.7$ {\AA}) hence permitting us
to identify the crossover from tunneling to contact with the
chemical crossover.

\begin{figure*}[t!]
  \centering
  \includegraphics[width=\textwidth]{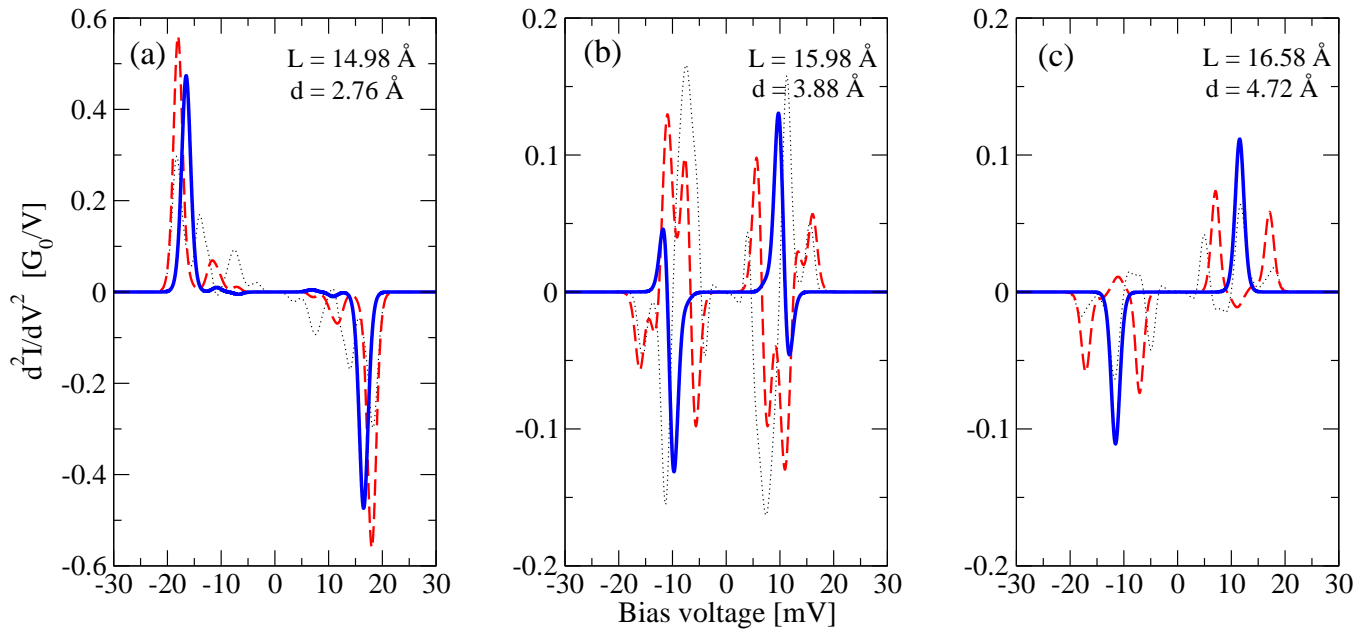}
  \caption{(Color online) Second derivative of the current versus bias voltage
  for three characteristic situations (a) contact, (b) crossover, and (c) tunneling.
 In each situation we consider different active vibrational regions: the two apex
 atoms only (thick blue line), the 10 pyramid atoms (thick dashed red curve),
 and both pyramids and first-layer atoms (dotted thin black curve).
 The signal broadening is due to temperature ($T=$ 4.2 K).}
  \label{fig:d2IdV2}
\end{figure*}


\section{Inelastic conductance}
\label{sec:InelasticConductance}

The behavior of the inelastic contributions to conductance is very
different in the two studied regimes. In the tunneling regime the
opening of the inelastic channel enhances the conductance of the
system, while the creation of a vibrational excitation in a high
conductance regime is a source of backscattering that decreases the
conductance. Figure \ref{fig:d2IdV2} shows the calculated change in
conductance (second derivative of the current with respect to bias
voltage $d^2 I/dV^2$) for the contact, crossover and tunneling
regions. These three typical cases---labeled (a), (b), and (c),
respectively---are indicated in the previous
Figs.~\ref{fig:EtotVsLength}--\ref{fig:TransVsApexDist} for easy
reference. We investigate how the inelastic conductance change
depends on how many atoms in the junction that are considered
active: in \Figref{fig:d2IdV2} the thick blue line is the spectrum
corresponding to only the two apex atoms vibrating, the dashed red
curve to the 10 pyramid atoms vibrating, and the dotted black curve
to the pyramids and first-layer atoms vibrating (42 atoms). In this
way we follow the convergence of the calculations as the vibrational
region is gradually enlarged. The essential data from these
calculations are summarized in \Tabref{tab:d2IdV2}.

From the simplest case when only the two apex atoms are vibrating,
we arrive at the conclusion that only the two longitudinal stretch
modes contribute to the change in conductance, leading to the
qualitative known result of increase of the conductance in tunneling
regime and decrease in contact. The crossover case
\Figref{fig:d2IdV2}(b) presents a combination of an increase in
conductance from the ABL mode and a decrease from the CM mode.

This behavior is a signature of the different processes of
conduction. In the tunneling case, the tunneling process is
determined by the more slowly-decaying components of the electron
wave function of the surface. Because of the exponential tunneling
probability dependence on distance a mode that modulates the
tunneling gap is expected to contribute positively to the
conductance.\cite{Lorente:2005} Indeed this is the case for the ABL
mode. Furthermore, the CM mode that correspond to a fixed apex-apex
distance cannot contribute positively, neither  the transverse modes
because none of them \emph{decrease} the apex-apex distance from the
equilibrium position during a vibration period. Instead, the CM mode
is found to contribute negatively to the conductance,
cf.~\Tabref{tab:d2IdV2}. A simplified model presented in the next
section will explain this observation.

In the contact case, the electronic structure responsible for the
conduction process is largely concentrated upon the apex atom, hence
the transport is being modified by the motion of basically only
these atoms. Indeed both the ABL and CM modes lead to drops in the
conductance as is evident from \Figref{fig:d2IdV2}(a) and
\Tabref{tab:d2IdV2}. The transverse modes give essentially no
signal; this is similar to the findings for atomic gold wires where
the transverse modes cannot couple because of
symmetry.\cite{Frederiksen:2004,Frederiksen:2006}

Figure~\ref{fig:d2IdV2} shows how the inelastic spectrum is modified
if we increase the vibrational region by allowing more atoms to
vibrate. In the tunneling and contact cases we see that the single
main peak splits up into a number of peaks, indicating that the apex
vibrations are actually coupled with the vibrations in the bulk.
For the contact case the broadening of the signals is expected to be
directly influenced by the phonon density of states of the bulk. As
was first shown by Yanson,\cite{Yanson:1974} the spectroscopy of
microcontacts at low temperatures---a technique nowadays referred to
as point contact spectroscopy---reveals a signal in $d^2 I/dV^2$
which is a direct measurement of the Eliashberg function
$\alpha^2F$, i.e., roughly speaking the product of the squared
electron-phonon coupling matrix element $\alpha$ and the phonon
density of states $F$, averaged over the Fermi
sphere.\cite{Jansen:1980} In the case of microcontacts the measured
signal is predominantly due to the transverse modes. This is in
contrast to our case of the atomic point contact, where we only find
signals from the longitudinal modes. However, from
\Figref{fig:d2IdV2}(a) we notice a signal broadening by increasing
the vibrational region, pointing towards the vibrational coupling to
the bulk modes.

In the crossover region between tunneling and contact,
\Figref{fig:d2IdV2}(b) shows a dramatic change depending on the size
of the vibrational region. Different modes give positive or negative
contributions in the conductance, but in such a way that they lead
to an overall absence of (or relatively small)  variation in the
conductance, cf.~\Tabref{tab:d2IdV2}.

Comparing the total change in conductance \mbox{$\Delta
G=G(V\gg\hbar\omega_\lambda)-G(V=0)$} induced by all modes (for the
tunneling, crossover, and contact situations), we find that the
calculations with different vibrational regions give almost the same
results. As found in \Tabref{tab:d2IdV2}, we thus conclude that to a
first approximation we can describe $\Delta G\approx \delta
G_\textrm{ABL}+\delta G_\textrm{CM}$, i.e., the overall conductance
change can be estimated with the minimal vibrational region (the two
apex atoms). This simple approach does however not accurately
describe details of the $d^2 I/dV^2$ spectrum.

\begin{table*}
\begin{tabular}{cccccccccccccccccc}
  \hline\hline
  L & d &\quad& $\tau$
    &\mbox{\quad}& $\omega_\textrm{ABL}$ & $\delta G_\textrm{ABL}/\mathrm{G}_0\tau$
    &\mbox{\quad}& $\omega_\textrm{CM}$ & $\delta G_\textrm{CM}/\mathrm{G}_0\tau$
    &\mbox{\quad}& $\Delta G/G$ &\mbox{\quad}& $\Delta G/G$&\mbox{\quad}& $\Delta G/G$\\
  {[\AA]} & [\AA] && - && [meV] & [\%] && [meV] & [\%] && [\%]\footnote{Only apex atoms vibrating, device includes first-layer atoms}
  && [\%]\footnote{Apex and base atoms vibrating, device includes first-layer atoms}
  && [\%]\footnote{Pyramids and first-layer atoms vibrating, device includes first- and second-layer atoms}\\
  \hline
  14.98 & 2.76 && 0.988 && 16.52 & -0.104 && 10.83 & -0.002 && -0.105 && -0.146 && -0.151 \\
  15.38 & 2.88 && 0.978 && 12.46 & -0.145 && 9.81 & -0.005 && -0.149 && -0.206 && --- \\
  15.78 & 3.23 && 0.857 && 7.57 & -0.223 && 9.73 & -0.014 && -0.235  && -0.340 && --- \\
  15.98 & 3.88 && 0.395 && 9.80 & 0.077 && 11.47 & -0.035 && 0.045 && -0.006 && -0.032 \\
  16.18 & 4.22 && 0.208 && 11.00 & 0.224 && 11.86 & -0.045 && 0.181 && 0.193 && --- \\
  16.58 & 4.72 && 0.063 && 11.60 & 0.430 && 12.04 & -0.053 && 0.377 && 0.395 && 0.332 \\
  \hline\hline
\end{tabular}
\caption{Characteristic data for the six structures representing the
evolution of the junction from tunneling to contact regimes. The
columns describe the electrode separation $L$, apex-apex distance
$d$, transmission $\tau$, vibrational energies $\hbar\omega_\lambda$
and mode-specific conductance changes $\delta G_\lambda$ (for the
ABL and CM modes), and the total conductance change $\Delta G/G$
from all modes (calculated for three different sizes of the
vibrational region).} \label{tab:d2IdV2}
\end{table*}


\section{Discussion}

The effect of the tunneling to contact crossover has important
implications in the inelastic conductance since in the first case
the inelastic effects trend to increase and in the second case to
diminish the electron conduction. From the results of the previous
section, we have seen that this crossover roughly takes place at the
same range of distances as in the elastic conductance case. By
looking at the transmission in the elastic conductance case, we
conclude that when the transmission is $\tau=1/2$ we should also be
near the crossover between tunneling to contact in the inelastic
case one. This finding is similar to the crossover found for the
single-state impurity model analyzed in
Ref.~[\onlinecite{Paulsson:2005}]. However, in the present case, the
system is not obviously modeled with a single-state impurity.
Instead we can easily reproduce the same kind of analysis for a
slightly more sophisticated model, where two impurities are
connected to reservoirs and interacts via a hopping term between
them. Under symmetric conditions this system is described by
\begin{eqnarray}
\mathbf{H} = \left[\begin{array}{cc}
    \varepsilon_0 & t \\
    t & \varepsilon_0
 \end{array}\right],\quad
\mathbf{\Gamma}_L = \left[\begin{array}{cc}
    \gamma & 0 \\
    0 & 0
\end{array}\right],\quad
\mathbf{\Gamma}_R = \left[\begin{array}{cc}
    0 & 0 \\
    0 & \gamma
\end{array}\right],
\end{eqnarray}
where the Hamiltonian $\mathbf{H}$ includes on-site energies
$\varepsilon_0$ and a hopping matrix element $t$. The level
broadening functions $\mathbf{\Gamma}_\alpha$ describes the coupling
of the sites to the contacts $\alpha=L,R$ with strength $\gamma$
(which in the wide band approximation are considered energy
independent). The corresponding retarded Green's function is
\begin{eqnarray}
\mathbf{G} &=& [\varepsilon_\textrm{F}\mathbf{1}-\mathbf{H}+i(\mathbf{\Gamma}_L+\mathbf{\Gamma}_R)/2]^{-1}\nonumber\\
&=& \frac{2}{(2\Delta\varepsilon + i \gamma)^2 - 4 t^2}\left[
\begin{array}{cc}
 2\Delta\varepsilon+i \gamma  & 2t \\
 2t & 2\Delta\varepsilon+ i\gamma
\end{array}
\right],\qquad
\end{eqnarray}
where in our case
$\Delta\varepsilon=\varepsilon_\textrm{F}-\varepsilon_0\ll\gamma$
holds since the level positions would be close to the Fermi energy
$\varepsilon_\textrm{F}$ (the on-resonance case). The transmission
becomes
\begin{eqnarray}
\tau =\frac{16 t^2 \gamma ^2}{\left(4 t^2+\gamma ^2\right)^2}
+\mathcal O(\Delta \varepsilon^2),
\end{eqnarray}
where perfect transmission $\tau=1$ is obtained for $t=\gamma/2$.

Inspired by our electron-phonon coupling matrices obtained from the
full DFT calculations, we assign the following forms to the
longitudinal ABL and CM mode couplings
\begin{eqnarray}
\mathbf{M}_\textrm{ABL} = \left[
\begin{array}{cc}
 m_1 & m_2 \\
 m_2 & m_1
\end{array}
\right],\quad
\mathbf{M}_\textrm{CM} = \left[
\begin{array}{cc}
 m_3 & 0 \\
 0 & -m_3
\end{array}
\right].
\end{eqnarray}
The ABL mode is symmetric and generally described by two coupling
strengths: $m_1$ represents an on-site modification via a change in
the electrode coupling, whereas $m_2$ is a modulation of the hopping
between the apexes. Correspondingly, the CM mode which is asymmetric
bears an asymmetric on-site modulation $m_3$ and no hopping
modulation (fixed apex-apex distance). With these expressions we can
simply evaluate \Eqref{eq:Conductance} to find the following
inelastic conductance changes
\begin{eqnarray}
\frac{\delta G_\textrm{ABL}}{\mathrm{G}_0\tau}
    &=& \frac{\left(16 t^4-24 \gamma ^2 t^2+\gamma ^4\right)}
        {\left(4 t^2+\gamma ^2\right)^2}\nonumber\\
    &&\times
    \left(\frac{{m_2}^2}{t^2} + \frac{16 m_1 m_2 \Delta \varepsilon}{4 t^3 +
    t{\gamma}^2}+\mathcal O(\Delta \varepsilon^2)\right),\\
\label{eq:CMcontrib} \frac{\delta G_\textrm{CM}}{\mathrm{G}_0\tau}
    &=& -\frac{16
        m_3^2 \gamma ^2}{\left(4 t^2+\gamma ^2\right)^2}+\mathcal O(\Delta \varepsilon^2),
\end{eqnarray}
We first discuss the conclusions to be drawn about the ABL mode.
Notice that $\delta G_\textrm{ABL}$ is only weakly dependent on the
the on-site coupling element $m_1$ and vanishes on resonance
($\Delta\varepsilon=0$). In the tunneling limit ($t\rightarrow 0$)
we find that
\begin{eqnarray}
\lim_{t\rightarrow 0}\frac{\delta
G_\textrm{ABL}}{\textrm{G}_0}=\frac{16 m_2^2}{{\gamma }^2}+\mathcal
O(\Delta \varepsilon^2),
\end{eqnarray}
i.e., the ABL mode gives a \emph{positive} contribution to the
conductance proportional to the square of coupling strength $m_2$.
In the contact limit ($\tau\rightarrow 1$) we find
\begin{eqnarray}
\lim_{\tau\rightarrow 1}\frac{\delta G_\textrm{ABL}}{\textrm{G}_0}=-
\frac{4m_2^2}{{\gamma }^2} - \frac{16 m_1m_2 \Delta
\varepsilon}{{\gamma }^3} +\mathcal O(\Delta \varepsilon^2),
\end{eqnarray}
i.e., the ABL mode gives here a \emph{negative} contribution. The
exact crossover between an increase and a decrease is determined by
solving $\delta G_\textrm{ABL}=0$, which indeed yields $\tau=1/2$ as
is the case for the single-site case.\cite{Paulsson:2005}

Next, we see from \Eqref{eq:CMcontrib} that the conductance change
$\delta G_\textrm{CM}$ from the CM mode is always negative (i.e.,
the CM mode backscatters even in the tunneling regime).

These results thus permit to rationalize the crossover from
tunneling to contact for the inelastic conductance---as found
numerically in \Secref{sec:InelasticConductance}---as taking place
around a transmission of $\tau=1/2$.


\section{Summary and conclusions}

The evolution of the inelastic signals from the tunneling to contact
regimes has been studied through DFT calculations. We have compared
our results with the crossover between bonding and rupture of the
atomic junction by studying the geometric and electronic structures
of the junction, together with the strength of the atomic bonds and
the associated vibrational frequencies. This permitted us to find a
typical transition distance between electrodes where a small change
leads to a large readjustment of the apex-apex atom distance, as
well as a change of the strength of interactions as revealed by the
total energy, the strain and the frequencies of the system's modes.

The conductance has been calculated by first evaluating the
transmission of electrons through the system, and second by
calculating the conductance change due to the excitation of
vibrations. As found in previous studies\cite{Paulsson:2005} the
change in conductance due to inelastic effects permits to
characterize the crossover from tunneling to contact. The most
notorious effect being a decrease of conductance in the contact
regime to an increase in the tunneling one when the bias voltage
exceeds the vibrational thresholds. Our DFT-based calculations show
that the effect of vibrational modes in the $d^2I/dV^2$ spectra is
rather complex, in particular when modes localized in the contact
region are permitted to extend into the electrodes. As an example,
we find that certain modes can give rise to decreases in conductance
when in the tunneling regime, opposite to the above mentioned
result. Whereas details in the inelastic spectrum depends
sensitively on the size of the vibrational region, we find that the
magnitude of the overall change in conductance can actually be
reasonably described with just the minimal case where only the apex
atoms vibrate. This means that while the modes are rather
delocalized the region of inelastic scattering is localized around
the apex atoms.

By comparing our results with a simplified model, we conclude that
in this single eigenchannel problem the tunneling to contact
crossover takes place exactly at $\tau=1/2$, in agreement with the
findings for the elastic conduction process and the chemical
crossover. Hence, we can trace back the origin of the conduction
process, both in the presence and absence of vibrational excitation,
to the same type of underlying electron structure that determine the
electrode's chemical interaction and the electron
conductance.\cite{Hofer:2003}

\acknowledgments{} The authors acknowledge many valuable discussions
with A.-P.~Jauho. This work, as a part of the European Science
Foundation EUROCORES Programme SASMEC, was partially supported by
funds from the SNF and the EC 6th Framework Programme. Computational
resources were provided by the Danish Center for Scientific
Computing (DCSC).


\bibliographystyle{apsrev}

\end{document}